%% file: make_all.tex
\documentclass{he_symp}
\usepackage{psfig,graphicx,epsfig}
\usepackage{color}
\usepackage{amsmath,amssymb,epic,eepic,array}
\begin{document}
\renewcommand{\FirstPageOfPaper }{ 300}\renewcommand{\LastPageOfPaper }{ 302}\include{he_symp_pacini}

\clearpage

\end{document}

%% file: he_symp_pacini.tex
%
%
\newcommand\simgt{\lower.5ex\hbox{$\; \buildrel > \over \sim \;$}}

\title{Neutron Stars, Pulsars and Supernova Remnants: concluding remarks}
\author{F. Pacini\inst{1,}\inst{2}}
\institute{Arcetri Astrophysical Observatory, L.go E. Fermi, 5, I-50125 Firenze,Italy
\and Dept. of Astronomy and Space Science, University of Florence, L.go E. Fermi, 2, I-50125 Firenze, Italy}
\maketitle

\section{Introduction}

More than 30 years have elapsed since the discovery of pulsars 
(Hewish {\it et al.} \cite{hew68}) and the realization that they are 
connected with rotating magnetized neutron stars (Gold \cite{gold68}; 
Pacini \cite{p67}, \cite{p68}). It became soon clear that these objects 
are responsible for the production of the relativistic wind observed in 
some Supernovae remnants such as the Crab Nebula.

For many years, the study of pulsars has been carried out mostly in the 
radio band. However, many recent results have come from observations at 
much higher frequencies (optical, X-rays, gamma rays).  These observations 
have been decisive in order to establish a realistic demography and have 
brought a better understanding of the relationship between neutron stars 
and SN remnants.

The Proceedings of this Conference cover many aspects of this relationship 
(see also previous Conference Proceedings such as Bandiera {\it et al.} 
\cite{elba98};  Slane and Gaensler, \cite{bost02}). Because of this reason, 
my summary will not review all the very interesting results which have been 
presented here and I shall address briefly just a few issues. The choice of 
these issues is largely personal: other colleagues may have made a different 
selection.

\section{Demography of Neutron Stars:  the role of the magnetic field}
For a long time it has been believed that only Crab-like remnants 
(plerions) contain a neutron star and that the typical field strength of 
neutron stars is $10^{12}$ Gauss. The basis of this belief was the lack 
of pulsars associated with shell-type remnants or other manifestations of 
a relativistic wind. The justification given is that some SN explosions 
may blow apart the entire star. Alternatively, the central object may 
become a black hole. However, the number of shell remnants greatly exceeds 
that of plerions: it becomes then difficult to invoke the formation of black 
holes, an event much more rare than the formation of neutron stars.

The suggestion that shell remnants such as Cas A could be associated with 
neutron stars which have rapidly lost their initial rotational energy because 
of an ultra-strong magnetic field  $B \sim 10^{14}-10^{15}$ Gauss  
(Cavaliere \& Pacini, \cite{cav70}) did receive little attention. The 
observational situation has now changed: a compact thermal X-ray 
source has been discovered close to the center of Cas A (Tananbaum, 
\cite{tanan99}) and it could be the predicted object. Similar sources have 
been found in association with other remnants and are likely to be neutron 
stars. We have also heard during this Conference that some shell-type 
remnants (including Cas A) show evidence for a weak non-thermal X-ray 
emission superimposed on the thermal one:  this may indicate the presence of 
a residual relativistic wind produced in the center.  Another important 
result has been the discovery of neutron stars with ultra-strong magnetic 
fields, up to $10^{14}-10^{15}$ G. In this case the total magnetic energy 
could be larger than the rotational energy ("magnetars"). This possibility 
had been suggested long time ago (Woltjer, \cite{woltj68}). It should be 
noticed, however, that the slowing down rate determines the strength of the 
field at the speed of light cylinder and that the usually quoted surface 
fields assume a dipolar geometry corresponding to a braking index $n=3$.  
Unfortunately the value of $n$  has been measured only in a few cases and 
it ranges between $1.4-2.8$ (Lyne {\it et al.}, \cite{lyne96}).

The present evidence indicates that neutron stars manifest themselves in 
different ways:
\begin{itemize}
\item{Classical radio pulsars (with or without emission at higher frequencies) 
where the rotation is the energy source.}
\item{Compact X-ray sources where the energy is supplied by accretion 
(products of the evolution in binary systems).}
\item{Compact X-ray sources due to the residual thermal emission from a hot 
surface.}
\item{Anomalous X-ray pulsars (AXP) with long periods and ultra strong fields 
(up to $10^{15}$ Gauss).  The power emitted by AXPs exceeds the energy loss 
inferred from the slowing down rate. It is possible that AXPs are associated 
with magnetized white dwarfs, rotating close to the shortest possible period 
($5-10$ s) or, alternatively, they could be neutron stars whose magnetic 
energy is dissipated by flares.
}
\item{Soft gamma-ray repeaters.}
\end{itemize}
In addition it is possible that some of the unidentified gamma ray sources 
are related to neutron stars.
The present picture solves some previous inconsistencies. For instance, 
the estimate for the rate of core-collapse Supernovae (roughly one every 
30-50 years) was about a factor of two larger than the birth-rate of 
radio pulsars, suggesting already that a large fraction of neutron stars 
does not appear as radio pulsars.

The observational evidence supports the notion of a large spread in the 
magnetic strength of neutron stars and the hypothesis that this spread is 
an important factor in determining the morphology of Supernova remnants. 
A very strong field would lead to the release of the bulk of the rotational 
energy during a short initial period (say, days up to a few years): at later 
times the remnant would appear as a shell-type. A more moderate field (say 
$10^{12}$ Gauss or so) would entail a long lasting energy loss and 
produce a plerion.

\section{Where are the pulses emitted ?}
Despite the great wealth of data available, there is no general consensus 
about the radiation mechanism for pulsars. The location of the region where 
the pulses are emitted is also controversial: it could be located close to the 
stellar surface or, alternatively, in the proximity of the speed of light 
cylinder.

The radio emission is certainly due to a coherent process because of the very 
high brightness temperatures ($T_b$ up to and above $10^{30}$ K have been 
observed). A possible model invokes the motion of bunches of charges sliding 
along the curved field lines with a relativistic Lorentz factor $\gamma$ such 
that the critical frequency $\nu_c \sim {c \over \rho} \gamma^3$ reaches or 
exceeds the radio band. Typically, this would entail  $\gamma \simgt 10^2$.

If we assume $T_b \sim 10^{30}$ K and  $\gamma \sim 10^2 -10^3$, the 
thermodynamical limitation $k T_b \sim m c^2 \gamma F$ requires a coherence 
factor $F$ (number of electrons in the bunch) of order  $\sim 10^{17}$.
At least one of the sizes of the bunch must be smaller than the emitted 
wavelength. The radio spectrum is determined by the distribution in size 
of the bunches and the effect of coherence is gradually lost at very high 
radio frequencies (Aloisio \& Blasi \cite{blasi02}). This may possibly 
explain the up-turn of the spectrum observed in some sources in the 
millimeter range, as reported here by Sieber.

Unfortunately the radio emission does not give sufficient information about 
the parameters of the source since many of them are affected by the degree 
of coherence. The situation is different if we consider pulsars which emit 
at higher frequencies (optical and X-ray bands), where the observed brightness
is compatible with an incoherent process.

A striking aspect of the optical pulses is the very strong dependency 
of the power upon the period (they are emitted only from a few sources).

A possible scenario assumes that the optical radiation is normal synchrotron 
radiation, emitted from particles which gyrate with a small pitch angle 
$\Psi$ around the field lines. Such a model, applied to the Crab pulsar, 
PSR 0531, can explain the observations with parameters compatible with those 
expected at the speed of light cylinder distance $R_L={c P \over 2 \pi}$:  
$\Psi \sim 10^{-2}$; $B_\perp \sim 10^4$ G; $\gamma \sim 10^2-10^3$.

The model leads to the expectation of a very fast decrease of the synchrotron 
intensity with period because of the combination of two factors: {\it a)} the 
reduced particles flux when the period increases; {\it b)} the reduced 
efficiency of synchrotron losses  (which  scale $\propto B^2 \propto R_L^{-6}
\propto P^{-6}$) at the speed of light cylinder (Pacini, \cite{P71};  
Pacini \& Salvati \cite{P83}, \cite{P87}). The prediction fits the observed 
secular decrease of the optical emission from the Crab Nebula and the 
magnitude of the Vela pulsar. A recent re-examination of all available 
optical data confirms that this model can account for the luminosity of the 
known optical pulsars (Shearer and Golden, \cite{shear01}).

If so, the optical radiation supports strongly the notion that the emitting 
region is located close to the speed of light cylinder.

\section{A speculation: can the thermal radiation from young neutron stars 
quench the relativistic wind?}

My final remarks concern the possible effect of the thermal radiation coming 
from the neutron star surface upon the acceleration of particles. This 
problem has been investigated for the near magnetosphere (Supper \& Truemper,
\cite{supp00}) and it has been found that the Inverse Compton Scattering (ICS) 
against the thermal photons is important only in marginal cases. However, if 
we assume that the acceleration of the relativistic wind and the radiation of 
pulses occur close to the speed of light cylinder, the situation becomes 
different and the ICS can dominate over synchrotron losses for a variety of 
parameters.

The basic reason is that the importance of ICS at the speed of light distance 
$R_L$ scales like the energy density of the thermal photons 
$u_\gamma \propto R_L^{-2} \propto P^{-2}$; on the other hand, the 
synchrotron losses are proportional to the magnetic energy density in the 
same region $u_B \propto R_L^{-6} \propto P^{-6}$.

Numerically, one finds that ICS losses dominate over synchrotron losses if 
$$ T_6 > 0.4 {B_{12}^{1/2} \over P_s}$$
where $T_6$ is the temperature in units $10^6$ K; $B_{12}={B \over 10^{12} {\rm G}}$; $P_s$ is the pulsar period in seconds).

The corresponding upper limit for the energy of the electrons, assuming that 
the acceleration takes place  for a length of order of the speed of light 
distance and that the gains are equal to the losses is given by:
$$ E_{\rm max} \simeq 1.2 \times 10^3 {T_6}^{-4}\ P_s\ {\rm GeV}.$$

Provided that the particles are accelerated and radiate in proximity of the 
speed of light cylinder distance, we conclude that the thermal photons can 
limit the acceleration of particles, especially in the case of young and 
hot neutron stars. It becomes tempting to speculate that this may postpone 
the beginning of the pulsar activity until the temperature of the star is 
sufficiently low. The main manifestation of neutron stars in this phase would 
be a flux  of high energy photons in the gamma-ray band, due to the 
interaction of the quenched wind with the thermal photons from the 
stellar surface. This model and its observational consequences are 
currently under investigation (Amato, Blasi, Pacini, work in progress).

